\documentclass[twocolumn,secnumarabic,amssymb, nobibnotes, aps, prd]{revtex4-1}

\usepackage{bm}
\usepackage{amsthm}
\usepackage{amsmath}
\usepackage[mathscr]{eucal}
\usepackage{xcolor,soul}
\usepackage{graphicx}
\usepackage{xcolor}

\begin{document}

% Use the \preprint command to place your local institutional report
% number in the upper righthand corner of the title page in preprint mode.
% Multiple \preprint commands are allowed.
% Use the 'preprintnumbers' class option to override journal defaults
% to display numbers if necessary
%\preprint{}

%Title of paper
\title{Rain Calms the Sea - The Impact of Entrained Air}

% repeat the \author .. \affiliation  etc. as needed
% \email, \thanks, \homepage, \altaffiliation all apply to the current
% author. Explanatory text should go in the []'s, actual e-mail
% address or url should go in the {}'s for \email and \homepage.
% Please use the appropriate macro foreach each type of information

% \affiliation command applies to all authors since the last
% \affiliation command. The \affiliation command should follow the
% other information
% \affiliation can be followed by \email, \homepage, \thanks as well.
\author{Juan M. Restrepo}\email[Corresponding Author: ]{restrepo@math.oregonstate.edu}
\affiliation{Department of Mathematics, Oregon State University, Corvallis OR 97331 USA}
\homepage[]{www.math.oregonstate.edu/~restrepo}
\author{Alex Ayet}
%\altaffiliation[Also at ]{LMD/IPSL, CNRS, Ecole Normale Sup\'erieure, PSL Research University, Paris 75005, France}
 \affiliation{
 Ifremer, University Brest, CNRS, IRD,  Laboratoire d'Oc\'eanographie Physique et Spatiale (LOPS), IUEM, Brest 29280, France}
\author{Luigi Cavaleri}
 \affiliation{ISMAR, Istituto di Scienze Marine,  
 Arsenale - Tesa 104, Castello 2737/F, 30122 Venice Italy}

%\thanks{}
%\altaffiliation{}

%Collaboration name if desired (requires use of superscriptaddress
%option in \documentclass). \noaffiliation is required (may also be
%used with the \author command).
%\collaboration can be followed by \email, \homepage, \thanks as well.
%\collaboration{}
%\noaffiliation

\date{\today}

\begin{abstract}
We propose a mechanism for the damping of short ocean gravity waves during rainstorms associated with  the injection of air bubbles 
by rain drops. The mechanism is proposed as one of the possible explanations that ascribe to rain 
a calming  effect on ocean surface waves.
A model is developed that shows how wave attenuation increases with the presence of air bubbles in the upper reaches of the ocean. The model
makes predictions of the effective wave dissipation coefficient, as a function of the volumetric ratio of air to water, as well as of
the rain rate. The model predicts dissipation rates that are in line with experimental estimates of the effective wave
damping rate. 
\end{abstract}
% insert abstract here

% insert suggested PACS numbers in braces on next line
\pacs{}
% insert suggested keywords - APS authors don't need to do this
%\keywords{}

%\maketitle must follow title, authors, abstract, \pacs, and \keywords
\maketitle

\section{Introduction}
\label{intro}

Mariners have known, for centuries, that  ``rain calms the sea." The sea surface roughness  can change  quickly, once  rain begins \cite{cavaleri2018sudden, cavaleri2000oceanographic}.
Several mechanisms   explaining the effect of rain on the short wind-waves have been proposed. In  \cite{le1990dynamic} the authors  
 conjectured that a momentum exchange between the rain drops and the sea takes place upon impact. Rain induces a fluctuating free-surface pressure as well as changes in the boundary shear stresses, and these changes lead to an increase in wave damping affecting, primarily,  the  high frequency spectral composition of the sea surface.
  In high wind conditions the authors suggest that these changes in the free boundary condition could lead to  damping as well as to an amplification of high frequency  wave components, leading to smoother or rougher sea states, respectively.  A model for impinging rain as a distributed pressure was proposed and investigated in \cite{veronmuessiens}. Within the irrotational framework they use, the rain pressure distribution modifies  the wave dispersion relation, affecting primarily the high frequency wave components. The authors also considered the effect of rain impact angle on the pressure distribution. They found that  waves are only damped for vertically-falling rain, but   can be amplified or damped otherwise.

 Several laboratory experiments have been  performed in which  
 the wave amplitude was measured prior and after going through
 a zone of rain. Linear damping rates \cite{lambbook}
 were then estimated. In this framework, the crucial parameter to be estimated  is $\nu_e$, the dissipation coefficient in the wave dissipation term (of the form  $\exp[-2 \nu_e k^2 t])$, where $t$  is time, and $k$ the wavenumber of the wave.
  In \cite{tsimplis} the dissipation rate was found to be  $3\times10^{-6}$ m$^2$s$^{-1}$, at rain rates of 140 mm h$^{-1}$ (recalculated in \cite{peirsonrain}). 
 The experiments in \cite{peirsonrain} yielded  a damping rate due to rain of  $8.5\times10^{-6}$ m$^2$s$^{-1}$, approximately. 
In \cite{harrisonveron} experiments were performed using freshwater rain on waves produced in a synthetic salt water tank. It was observed that the rain drops cratered the free surface, created a  stratified water column with sharp density gradients, and generated near-surface turbulence. 
 The   depth-integrated kinematic viscosity estimate, derived from these experiments,  was about ten times larger than the value reported in \cite{peirsonrain}. However, as pointed out by the authors,  there is a complex relation between the  depth-integrated kinematic viscosity, 
based upon a kinetic energy dissipation rate, and wave damping that affects comparative interpretations of their results with the older experiments.

Here we propose an alternative, but by no means  exclusive, mechanism for high frequency damping: the entrainment of air bubbles
by the impact of  rain drops.
 Air bubble creation due to impinging water drops and rain is well known (see
\citep{liow2007bubble} for a review. See also \citep{eggers2004}). 
A (heavy) rain  rate of the order of 100 mm h$^{-1}$ produces a rainwater flux of $3 \times10^{-6}$ m$^3$ m$^{-2}$ s$^{-1}$. Assuming 4 mm diameter rain drops (volume $3\times 10^{-8} $m$^3$), this translates into  100 drops per m$^{2}$ ({\it i.e.},  1 per $10\times10$ cm$^2$ s$^{-1}$).  In \cite{woolf} Woolf  suggests that under these conditions air injected into the sea can be a few cm$^3$ m$^{-2}$ s$^{-1}$.
In Section \ref{sec:model}, we show how the entrainment of air leads to an increase in the effective viscosity of the upper ocean. We then use results from  ocean acoustics (see \cite{oguz1990bubble},\cite{prosperetti1993impact}) to compute the distribution of air bubbles for a given rain rate (Section \ref{air}). How  changes in the effective viscosity affect gravity waves
 is taken up  in Section \ref{sec:waves}. Conclusions and a brief discussion appear in  Section \ref{sec:conclusions}.

\section{The Dynamics - During Rainfall}
\label{sec:model}

To determine the impact of  rain-induced air bubbles on wave attenuation, we perform homogenization on the Navier-Stokes equations. We assume that air bubbles are distributed uniformly  within the upper ocean. We seek to find the effective viscosity of the fluid averaged over a cell $\Omega$, of size $\ell^3$ (sub-wave scale), over which the distributions of the density and viscosity of the combined water and air bubble mixture are statistically stationary at wave scales.

Velocity and position are denoted by ${\bf u}=(u,v, w)$, ${\bf r}=(x,y, z)$, respectively. The free surface is denoted by $\eta$ and  $z = 0$
corresponds to the quiescent sea level. Gravity $g {\bf \hat k}$ points upwards, along increasing $z$. Velocity is scaled by $u_0$, 
the typical wave phase speed, space by $\ell$, time by $u_0/\ell$, density by 
$\rho_w$, the density of water, pressure by $\rho_w u_0^2$. We then define a Reynolds number $\alpha = u_0 \ell/\nu_w$, where $\nu_w$ is the viscosity of water. We also define a Froude number $1/\sqrt{\gamma} = u_0/\sqrt{g \ell}$, and a scale parameter $\epsilon \ll 1$ that represents the ratio (of spatial scales) of the microscale to the large scale. 
The scaling leads to
\begin{eqnarray}
&\alpha {\bf u}_t + \alpha {\bf u} \cdot {\bm \nabla} {\bf u} = - \alpha\gamma{\bm \nabla} \Pi+ {\bm \nabla} \cdot [D({\bf r} ){\bm \nabla} {\bf u}] \\
&{\bm \nabla} \cdot {\bf u} =0.
\label{NS}
\end{eqnarray}

 Let ${\bf R}=(X,Y, Z)$ be the large-scale space variable, such that ${\bm \nabla} = {\bm \nabla} + \epsilon {\bm \nabla}_{\bf R}$, and assume slow time $\partial_T= \delta \partial_t$. Also, assume that  $\Pi ({\bf r}, {\bf R}, T) = p- z + \epsilon^2 p_0 + \epsilon^4 p_1 + ...$, ${\bf u} = \epsilon ( {\bf u}_0 + \epsilon {\bf u}_1  + \epsilon^2 {\bf u}_2 ...)$ and 
$\eta = \alpha ( \eta_0 + \epsilon \eta_1  + \epsilon^2 \eta_2 ...)$.  
The orders are  $\alpha =  \mathcal{O}(\epsilon^2)$, $\delta = \mathcal{O}(\epsilon)$, and $\gamma =  \mathcal{O}(\epsilon^{-1})$.

In the absence of rain,    the dimensional $D({\bf r})$ is equal to $\nu_w$,  the kinematic viscosity of  water and the dissipative term in the momentum balance is equal to  $ {\bm \nabla} \cdot \nu_w {\bm \nabla} {\bf u}$. However, when rain is present and intense enough, the dissipative term becomes 
equal to $ {\bm \nabla} \cdot K {\bm \nabla} {\bf u}$, where $K$ is  an effective 
viscosity. This change in the momentum balance, it is proposed, comes about by the injection of  gas bubbles by rain in the near-surface ocean. 
Specifically, the gas voids in the ocean are created by the impact of the rain drops on the ocean surface. 
If the rain is intense enough, the voids can cause a significant change in viscous forces, within the momentum balance. 
Under the assumption that the subsurface air bubble distribution is  spatially homogeneous we can then propose a periodic arrangement of sub-wave cells.

\begin{figure}
\includegraphics[height=1.75in,width=2.7in]{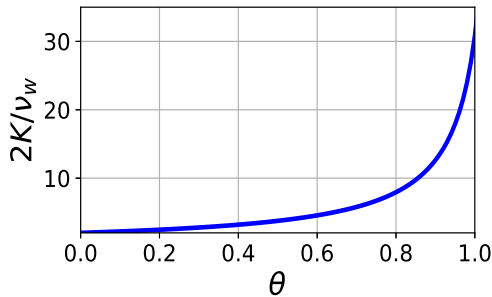}
\caption{Relative effective dissipation  $2 K/\nu_w$, as a function of volumetric ratio of air to water $\Theta$, using \eqref{Keq} .}
\label{fig:damping}
\end{figure}

Collecting by orders in $\epsilon$,   the momentum equation is, 
\begin{itemize}
\item $\mathcal{O}(\epsilon)$:
\[
{\bm \nabla} \cdot [D({\bf r} ){\bm \nabla} {\bf u}_0] = 0.
\]
Upon integrating and using periodicity it is clear that 
 ${\bf u}_0 = {\bf u}_0({\bf R},T)$.
\item $\mathcal{O}( \epsilon^2)$:
\begin{equation}
{\bm \nabla} \cdot [D({\bf r} ){\bm \nabla} {\bf u}_1] + {\bm \nabla} \cdot [D({\bf r} ){\bm \nabla}_{\bf R} {\bf u}_0]+ {\bm \nabla}_{\bf R} \cdot [D({\bf r} ){\bm \nabla} {\bf u}_0] = 0.
\label{eqorder1}
\end{equation}
Integrating twice over the cell $\Omega$, using the divergence theorem,  and using periodicity we obtain
\begin{equation}
0 = - \Omega  {\bm \nabla}_{\bf R}{\bf u}_0 + \int_\Omega \frac{d {\bf r}}{D({\bf r})} {\bm C_1} ({\bf R}).
\end{equation}
Hence, the tensor
\[
{\bf C}_1({\bf R}) = \left[ \frac{1}{\Omega} \int_\Omega \frac{d {\bf r}}{ D({\bf r})}  \right]^{-1} {\bm \nabla}_{\bf R} {\bf u}_0.
\]
Let 
\begin{equation}
K= \left[ \frac{1}{\Omega} \int_\Omega \frac{d {\bf r}}{ D({\bf r})}  \right]^{-1}.
\label{disp}
\end{equation}
Returning to (\ref{eqorder1}),
\begin{equation}
{\bm \nabla} {\bf u}_1 = -{\bm \nabla}_{\bf R} {\bf u}_0 + \frac{K}{D({\bf r})}{\bm \nabla}_{\bf R} {\bf u}_0,
\label{eq1}
\end{equation}
thus, 
\begin{equation}
{\bm \nabla}_{\bf R} \cdot( D {\bm \nabla} {\bf u}_1) = -{\bm \nabla}_{\bf R} \cdot (D {\bm \nabla}_{\bf R} {\bf u}_0) + {\bm \nabla}_{\bf R} \cdot (K {\bm \nabla}_{\bf R} {\bf u}_0).
\label{eq1a}
\end{equation}

\item $\mathcal{O}( \epsilon^3)$: the momentum balance reads
\begin{align}
\frac{\partial {\bf u}_0}{\partial T} + {\bf u}_0 \cdot {\bm \nabla}_{\bf R} {\bf u}_0 + {\bf u}_0 \cdot {\bm \nabla} {\bf u}_1 
 +  {\bm \nabla}_{\bf R} p_0 =  \nonumber \\
+ {\bm \nabla} \cdot [D({\bf r} ){\bm \nabla}_{\bf R} {\bf u}_1] 
 + {\bm \nabla}_{\bf R} \cdot [D({\bf r} ){\bm \nabla}_{\bf R} {\bf u}_0] \nonumber \\
+ {\bm \nabla} \cdot[D({\bf r} ){\bm \nabla} {\bf u}_2]
+ {\bm \nabla}_{\bf R} \cdot [D({\bf r} ){\bm \nabla} {\bf u}_1].
\label{eq3}
\end{align}
\end{itemize}
Using (\ref{eq1a}) in the last term in (\ref{eq3}) and averaging all quantities over  $\Omega$, leads to
\begin{equation}
\frac{\partial {\bf u}_0}{\partial T} + {\bf u}_0 \cdot {\bm \nabla}_{\bf R} {\bf u}_0 
 +  {\bm \nabla}_{\bf R} p_0 =    
 {\bm \nabla}_{\bf R}  \cdot [K {\bm \nabla}_{\bf R} {\bf u}_0]. 
\label{momeq}
\end{equation}
The homogenized incompressibility condition is
\begin{equation}
{\bm \nabla}_{\bf R} \cdot {\bf u}_0 = 0.
\label{conteq}
\end{equation}
The value of $K$ is equal to $\nu_w$, prior to a rainstorm, and it becomes the enhanced value,
\begin{equation}
K = \frac{\nu_w}{1 - \Theta \left(1 - \frac{\nu_w}{\nu_a} \right)},
\label{Keq}
\end{equation}
computed via (\ref{disp}),
where $\nu_w$  is the kinematic viscosity of water and $\nu_a$, its counterpart for air. The volumetric ratio of air to water is $\Theta$.
Air is about 100 times less viscous, hence, the viscosity of water with air bubbles produces a larger effective viscosity K, than the viscosity of water.
(If the ocean flow is turbulent, the ocean viscosity would be better described by the eddy viscosity, which can be substantially larger than the 
molecular viscosity of water). Figure \ref{fig:damping} describes how the effective dissipation coefficient $K$ changes, as a function of the volumetric ratio $\Theta$.

\section{Air Entrainment due to Rain}
\label{air}

\subsection{Rain Distribution}

The density of rain drops is assumed to follow the Marshall-Palmer distribution \cite{marshall1948distribution}.
 The density $N_r$ of rain drops of radius $r$ [mm] per unit volume  [number m$^{-3}$ mm$^{-1}$] is
\begin{equation}
    N_r(r, R) = N_0 \exp(-\Lambda(R) r),
\end{equation}
where $N_0 = 8 \times 10^3$ m$^{-3}$ mm$^{-1}$, $\Lambda(R) :=  8.2 R^{-0.21}$ [mm$^{-1}$].
where  $R$ [mm h$^{-1}$] is the rain rate.
Using $N_r$ we can then define 
the drop rate density (DRD), which describes the rate of falling rain drops of radius $r$  per surface area [number m$^{-2}$ s$^{-1}$ mm$^{-1}$]. Namely, 
\begin{equation}	
\text{DRD}(r, R) = w_a(r) N_r(r, R) ,
\end{equation}
where $w_a$ is the terminal velocity of drops in the air, approximated as
\begin{equation}
w_a \sim 4.6 \sqrt{2 r }	.
\end{equation}

\subsection{Bubble Production}

Rain drops generate subsurface air bubbles in the neighborhood of the sea surface (see \citep{prosperetti1993impact} for a review). 
Small rain drops produce small air bubbles with a very narrow distribution of radii.  Intermediate rain drops, with a radius between 0.55 - 1.1 mm, were shown not to produce air bubbles.  \cite{medwin1992anatomy}. For these the impinging rain drops do not have the kinetic energy necessary to produce the requisite conical crater and jetting of the sea surface that engulfs air. Large rain drops, with radii larger than 1.1 mm, 
create  a crater and a canopy on the sea surface, which by collapsing produces  a downward  liquid jet at the bottom of the crater, followed by the generation of an air  bubble.  At high rain rates the larger rain drops are responsible for the bulk of the gas injection, creating bubbles with a varied distribution of radii. 

The production of air bubbles  by rain can be measured by acoustic means \cite{prosperetti1993impact}. Oguz and Prosperetti \cite{oguz1990bubble} classify air bubble production by falling rain drops in  two regimes.  Small rain drops, of radii between 0.41 and 0.55 mm, create  {\it type I} air bubbles. These have  a radius of $0.22$ mm and are created at the bottom of a conical crater created on the ocean surface by the impinging rain drop. These air bubbles have a narrow acoustic spectrum with a spectral peak at  $14$ Hz. The acoustic spectrum for these type I air bubbles was found to be insensitive to the rain rate.   (Medwin {\it et al.}  \cite{medwin1990impact} observed that when  rain hits the ocean surface at an angle, owing to strong winds or very steep wave conditions,  the acoustic  spectrum peak shifts downward and  there is a broadening of the spectrum at higher frequencies).   The type I air  bubbles do not contribute significantly to total submerged gas volume  and their contribution will be  ignored, in what follows.

 Rain drops with a  radius  greater than  1.1 mm   produce  {\it type II} air resonant bubbles of varying radius.  
The relationship between the rain drop radius $r$ [mm] and the peak acoustic emission frequency $f_0$ due to trapped air bubbles  was empirically determined  \cite{medwin1992anatomy} as
\begin{equation}
f_0 = \frac{160}{8 r^3} + 0.6.
\label{f0}
\end{equation}
The approximate relation between the near surface air bubble radius $a$ [mm]  and the peak acoustic emission is
\citep{medwin1997fundamentals}:
\begin{equation}
f_0 \sim 3.25 \times 10^3 / a.
\label{f02}
\end{equation}

The relation between the entrained air bubble radius $a(r)$ and the incident rain drop radius $r$, for large rain drops, thus reads
\begin{equation}
a(r) = 3.25 \times 10^3 \left(\frac{160}{8 r^3} + 0.6\right)^{-1}, \text{ for  $r > 1.1$.}
\end{equation}
%\begin{equation}
%f(r) = \left\{\begin{aligned}
%& 0.22 &\text{for } r \in (0.41, 0.55) \\
%&0 &\text{for } r \in (0.55,1.1)\\
%& 3.25 \times 10^3 \left(\frac{160}{8 r^3} + 0.6\right)^{-1} & %\text{for } r > 1.1
%\end{aligned}\right.
%\end{equation}

\begin{figure}
\includegraphics[scale=0.5]{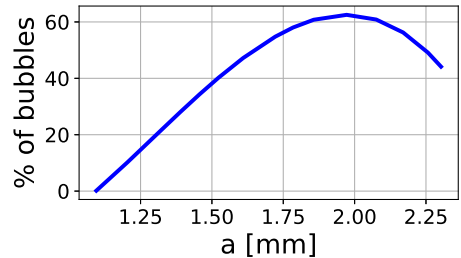}
\caption{Percentage of air bubbles produced by large rain drops  as a function of rain drop radius $r$ (from \cite{medwin1992anatomy}).}
\label{fig:production_rate}
\end{figure}

Small rain drops ($r < 0.55$ mm )  almost always produce type I air bubbles. 
However, this is not the case for the  larger rain drops. The distribution (\%) of the number of air bubbles produced, as a function of the rain drop radius $r$, was found  
 by Medwin {\it et al.},   \cite{medwin1992anatomy},  and is depicted in Figure \ref{fig:production_rate}. 

\subsection{Bubble Distribution in the Surface Boundary Layer}

We denote with  $P_r(r)$  the probability of production of an air bubble by a rain drop of radius $r$ (see Figure \ref{fig:production_rate}).  We also define 
the rate of air bubbles produced by rain drops of radius $r$ [number m$^{-2}$ s$^{-1}$ mm$^{-1}$] as
\begin{equation}
Q_a(r) =(\text{DRD} \times \text{P}_r) (r).
\end{equation}

The distribution of bubbles in the water is a balance between  (i) the production of bubbles by rain ($Q_a$), (ii) a sink term  that captures the loss of air bubbles that burst at the surface, (iii) vertical advection, (iv) attenuation of bubbles by diffusion. See \cite{woolf1991bubbles} and references therein. 

It is  assumed that the injection of bubbles occurs at a fixed depth, near the sea surface, 
that the bubble distribution is vertically homogeneous  (which cancels advection), and that diffusion takes place on longer times than the characteristic time of bursting of a bubble (the typical time of diffusion is 100 times longer than the upward motion). Armed with these assumptions  the following balance equation 
\begin{equation}
Q_a = w_w N(r,R)
\end{equation}
is obtained, 
where
\begin{equation}
w_w(a) \sim \frac{2}{9} \frac{g \rho_w (a \times 10^-3)^2}{\mu_w}
\end{equation}
is the terminal velocity of a bubble in water of dynamic viscosity $\mu_w$  \citep{batchelor1967introduction}, and $N$ is the density of bubbles produced by drops of radius $r$, per unit volume [number m$^{-3}$ m$^{-1}$]. Non-stationary conditions descriptions are also possible, by including the mixing due to wave turbulence (see \cite{oil1,massexchange}). 

The desired volumetric ratio of air to water is obtained as an integration of the density of bubbles, times their volume (assuming a spherical bubble), over the distribution of rain drops, {\it viz.}, 
\begin{equation}
    \Theta(R) = \int_{1.1}^{2.3} \frac{4 \pi }{3} [a(r) \times 10^{-3}]^3 \frac{(DRD \times P_r)(r)}{w_w[a(r)]} dr,
\end{equation}
where the limits of integration encompass  the volumetric contributions of type II bubbles.
%\subsection{Time evolution of the distribution}
%
%Here we suggest a mechanism for the evolution of the bubble distribution when rain stops.
%
%As mentioned above, the bubble diffusion processes are 100 times slower than the bubble upward motion. Thus, when rain stops, the bubble distribution decays due to the upward motion of the bubbles only.
%
%The decay of the bubble distribution can be determined from the knowledge of the bubble injection dept $h_i$ (\textbf{find it in the litterature?}). From that, one can assume that the bubble distribution for a radius $r$ is a step function: $N(r,R)$ above $h_i$ and zero below. When  rain stops, the location of the deepest bubble $h_i$ moves upward with a velocity $w_w(r)$. This sets the dyamics of evolution of the distribution $N(r)$ as a function of depth.

\section{The Effect of Rain on Gravity Waves}
\label{sec:waves}

\begin{figure} 
\includegraphics[scale=0.5]{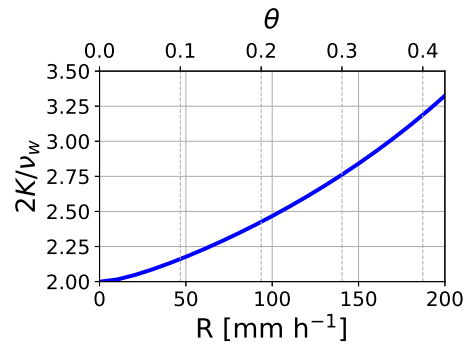}
\caption{Relative effective wave dissipation $2 K/\nu_w$, as a function of rain rate $R$ (bottom axis), as a function of volumetric ratio (top axis).  }
\label{fig:effective_density}
\end{figure}

The derivation of infinitesimal gravity waves dynamics, starting from Navier Stokes,  appears  in  \cite{lambbook}, Article 349, hence we will be brief.
We revert back to dimensional quantities in what follows. 
 The solution ${\bf u}_0$ satisfies the linear version of (\ref{momeq}) 
and (\ref{conteq}) and
the two stress conditions at the surface. Namely, 
\begin{align}
K [ \partial_X v_0 + \partial _Y u_0] &=0, \quad \text{at $z=0$}, \label{eqstress1} \\
-\partial _T \phi + (g - \frac{\tau }{\rho_w} \partial_{XX}) \eta_0 + 2 K \partial_Y v_0&=0,\quad \text{at $z=0$},
 \label{eqstress2}
 \end{align}
where   $\phi$ is the velocity potential, $\tau$ is the surface tension at the atmosphere/water interface and $\rho_w$ is the density of water. 
The solution of the system, assuming a vanishing velocity at depth,  is 
\begin{align}
\phi&=A e^{ k Z + i (k X -\sigma T)} \exp[-2 K k^2 T], \nonumber \\
\psi &=\frac{2 K  k^2  A}{\sigma} e^{kZ+ i (k X -\sigma T) } \exp[-2 K k^2 T ], \nonumber \\
\eta &= \frac{k A}{\sigma} e^{  i( k X -\sigma T)}  \exp[-2 K k^2 T ],
\label{longwave}
\end{align}
where $\psi$ is the streamfunction and $\sigma$ the angular frequency.  The solution (\ref{longwave}) represents infinitesimal progressive waves traveling 
in the $X$ direction. The dispersion relation for the waves is 
\[
\sigma^2 = g k + \tau' k^3,
\]
 where  $\tau'$ is the effective  surface tension. The effective wave dissipation $2K$  appears in the exponential factor
 $\exp[-2 K k^2 T] $. When the rainstorm is sufficiently intense, the wave dissipation  $2K$ will increase from
 $2 \nu_w$ to a higher value, depending on how much air is injected by the impinging rain drops (see Figure \ref{fig:damping}). 
 The dependence of the  effective wave dissipation  on the volumetric ratio and on the rain rate, via (\ref{Keq}), is depicted in  Figure \ref{fig:effective_density}.
(The rain rate range in  \cite{tsimplis} corresponds to a volumetric range 0-0.4).
As can be surmised from (\ref{longwave}), increases in the wave dissipation $2K$ leads to greater wave attenuation. The attenuation is most prominent in  the high frequency components 
of the wave spectrum due to the dispersive nature of the waves.

\section{Discussion and Conclusions}
\label{sec:conclusions}
The mechanism whereby rough seas may be made smoother when exposed to rainfall is not fully understood. Several conjectures have been proposed. These range from distributed pressure fluxes to changes in the shear in the neighborhood of the sea surface, induced micro-turbulence and fluctuations in buoyancy forces. Here we add another possible mechanism, namely, the injection of air bubbles by rain which then increases 
gravity wave damping.  The  model for how this  occurs  relies on homogeneization theory, and on well established theory and observations of 
the process whereby rain drops inject air into the ocean. The model describes the connection between rain rates, air bubble distribution and wave damping rates. 
Our analysis does not  take into account changes in the structure of the turbulent boundary layer by the presence of the air bubbles. Our model could be  extended to  incorporate vertical structure under commonly-used adiabatic assumptions for boundary layers.   
Future improvements on the model would account for  the effect of wind,  rain impact direction, in addition to bubble distribution  heterogeneity.

Experiments using very high rain rates suggest that  the effective dissipation can increase 3-10 times when rain is present, as compared to without rain. Our model exhibits increases of about half of that. A distinguishing characteristic of the proposed mechanism is that it can  only dampen high frequency waves. Alternative mechanisms that connect rain to fluctuations in the shear and pressure gradients, however, can dampen as well as increase the presence of high frequency ocean waves under some situations.
The possibility that several agents are responsible for 
the phenomenon of calming seas by rain  cannot be discounted without further  analysis and experiments.
\begin{acknowledgments}
The authors wish to thank the Kavli Institute of Theoretical Physics (KITP), at the University of California, Santa Barbara, for their hospitality and for supporting this research  project. The KITP is supported in part by the National Science Foundation under Grant No. NSF PHY-1748958. AA is supported by DGA grant No. D0456JE075 and the French Brittany Regional Council. LC has  been partly supported by  EU contract 730030 call H2020-EO-2016 'CEASELESS.'  JMR was supported by grant NSF/OCE \#1434198.
\end{acknowledgments}

% Create the reference section using BibTeX:
\bibliography{swwe}

\end{document}